# Spatial Templates for Studies of Patterns and Interactions in Dielectric-Barrier Discharges

S. Alsum, M. Carter, A. Hess, J. Lampen, and M. Walhout*
*Department of Physics & Astronomy, Calvin College, Grand Rapids, MI 49546, USA*

We use various electrode layouts to tailor the profile of the electric field in a two-dimensional (2D) dielectric-barrier-discharge system, thereby creating spatial templates for the lateral distribution of discharge filaments. Our discussion focuses on the formation of patterns and the interactions between filaments in several template configurations. Hexagonal and square lattice templates produce patterns reminiscent of the surface phases of adatoms on crystal surfaces. For the square lattice, a lateral displacement between opposing lattice electrodes modifies the pattern-forming behavior of the system. In a stripe-electrode arrangement, each discharge filament terminates in two dissimilar footprints, and the resulting pattern exhibits smectic behavior. A circular ring template establishes periodic boundary conditions and produces a new pattern in which adjacent filament footprints differ in shape. Finally, electrodes featuring isolated and twinned pins are used to investigate the pairing and clustering of filaments. We find that under certain conditions, filament pairs may stabilize at either of two characteristic bonding distances. We also study the range of distances over which a filament will either prevent the ignition of or annihilate nearby filaments.

## I. INTRODUCTION

Dielectric-barrier discharges (DBDs) have played important roles in applied physics and in the basic science of spatiotemporal pattern-formation [1, 2]. Investigations of DBD-based metamaterials have bridged these two lines of research, suggesting that discharge patterns themselves may lead to new applications [3]. Our recent paper demonstrated the use of a pin-grid electrode that produced a two-dimensional (2D) lattice potential, which served as a spatial template for the formation and selection of patterns [4]. That work made several thematic connections between DBDs and other physical systems in which patterns form on substrate lattices. It also introduced the idea that tailored electrodes can be put to use in studies of the interactions between filamentary microdischarges. In the present paper we expand and improve on our earlier experiments by creating spatial templates out of transparent electrodes of various shapes. We describe preliminary explorations of patterns and interactions in several electrode configurations.

To begin, we review some general characteristics of a DBD system such as ours. We consider a device in which a driving voltage $V = V_0 \sin(2\pi\nu t)$ is applied across a "sandwich" composed of two dielectric slabs and an interposed layer of gas. For now we assume that the slabs are flat, or "featureless," and that the device operates near atmospheric pressure and in a frequency range of 1 kHz < $\nu$ < 100 kHz. We refer to the width of the gas layer as the "gap width" and use the variable $z_0$ to represent its value. If $V_0$ is sufficiently high, brief discharges occur in the gap as the voltage either rises or falls in successive half-cycles of the driving oscillation.

The defining characteristic of this (or any) DBD system is the capacity of the dielectric barriers to intercept charged particles that are accelerated through the gas toward the conducting electrodes. The charges that accumulate on the dielectric surfaces have the effect of attenuating the electric field in their vicinity. Just a few nanoseconds of accumulation can be enough to weaken the local field so severely that the discharge is extinguished. This mechanism is responsible for the pulsed nature DBD systems.

Most often, devices like ours operate in a filamentary mode: when the gas breaks down, the discharge takes the form of narrow microdischarges, or filaments, which strike directly across the gap and leave localized patches of charge on the dielectrics. Multiple filaments may form simultaneously at various positions in the lateral dimensions of the system. Once a filament strikes at a specific location, that same location becomes a likely site for another filament-ignition during the subsequent half-cycle of the driving oscillation. This repetitive tendency is due to the "memory effect" of localized ignition residues—that is, charged and excited species that can survive for many microseconds and enhance the local probability of breakdown when the field is reversed [5, 6, 7]. For an imaging system that averages over many cycles of the driving voltage, this effect produces luminous, quasi-stationary filaments that are stable over long time scales (much longer than $1/\nu$).

Pattern formation within the spatial distributions of DBD filaments has been investigated extensively for the case of flat, featureless electrodes in both 1D and 2D geometries [8, 9]. Experimental studies have benefitted greatly from the development of imaging techniques that offer both spatial and temporal resolution [10, 11, 12]. These techniques have shown conclusively that many patterns that can be observed with the naked eye are actually time-averaged superpositions of multiple sub-patterns. In such cases, a composite image arises from two or more "discharge stages" that strike at distinct temporal phases of each half-cycle of the driving oscillation.

Experimenters have also invented techniques for measuring the surface-charge distributions, or "charge footprints," deposited by filaments [13]. Such work will be relevant in our discussion of filament interactions. However, we must note that our imaging system records only "optical footprints," which coincide with the flared regions of luminous gas that appear at the ends of filaments. While it is not unreasonable to expect that an optical footprint and its corresponding charge footprint will have similar shapes, it has been shown that the surface charge can spread laterally beyond the visible perimeter of the optical footprint [14]. Further investigations of the correlations between charge footprints and optical footprints may provide answers to certain questions arising from the studies described below.

## II. EXPERIMENTAL SETUP

The aim here is to use electrodes of various shapes to create spatial templates for DBD filament patterns. The experimental system used in [4] enabled us to meet this goal in important ways, but it was limited by the fact that the pin-grid electrode was metallic and therefore opaque. The second electrode consisted of a flat layer of water and served as a viewing window. This configuration resulted in a "single-sided" template that produced asymmetries between opposite ends of each discharge filament. For the present paper we have developed a new experimental system in which these asymmetries can be modified or even eliminated.

The main innovation in our new setup is the use of two water electrodes that follow the contours of shaped dielectric barriers. Prior to being installed in the system, the barriers start out as featureless plastic slabs (transparent acrylic or polycarbonate), but by machining divots or grooves into them we create "negative features" that become points or ridges in the profiles of the water electrodes. As indicated in Fig. 1, each water electrode is contained inside a plastic cartridge, one face of which is the intended dielectric barrier. This approach allows us to explore a variety of electrode combinations and to create either single- or double-sided templates.

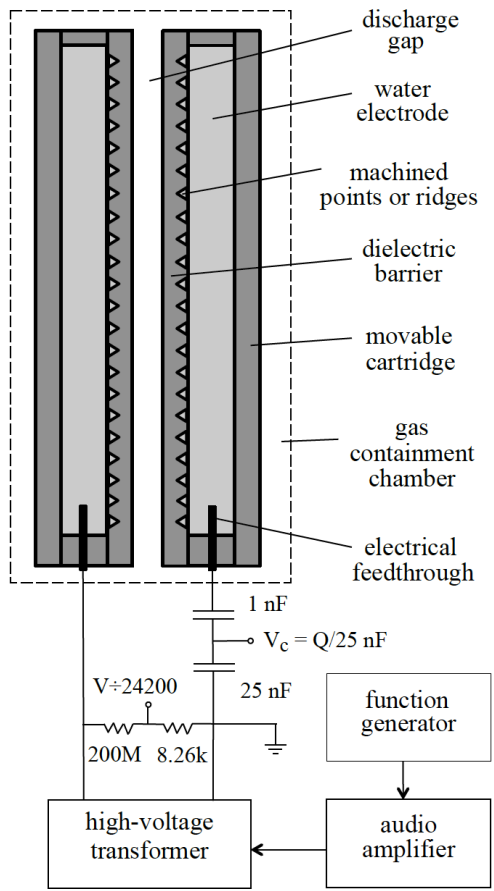

FIG. 1. Schematic of experimental setup.

All of the electrode features employed in this paper—whether pins or ridges—are machined into the dielectrics with a cone-shaped milling bit. The bit has an opening angle of 60° and a slightly rounded tip with radius 0.13 mm. Our dielectric slabs are 3.0(1) mm in thickness, and all of our cuts leave 0.5(1) mm of material between the tip of the bit and the opposite side of the dielectric barrier. We can estimate the local field amplitude in the gas gap using the approximation $E_0 = V_0/[z_0 + (\delta_1 + \delta_2)/\kappa]$, where $z_0$ is the width of the gas gap, $\delta_1$ and $\delta_2$ are the local thicknesses of the opposed dielectric barriers, and $\kappa \approx 3$ is the barrier's dielectric constant. This estimate neglects the geometric field enhancement that arises from the sharpness of the electrode pins and ridges. For a typical double-sided template with $V_0 = 1000$ V and $z_0 = 2$ mm, $E_0$ varies from ~ 250 V/mm in the featureless regions to > 400 V/mm at points between opposing pins or ridges.

Further details of the system's construction are as follows. The water electrodes measure 0.9 cm in thickness and 10 cm x 10 cm in the lateral directions. Each electrode cartridge consists of front and back plates (either of which may serve as the dielectric barrier) glued onto a central polycarbonate frame. The outside edges of the frame are drilled to make holes for mounting screws, electrical connections, and fill/drain ports. Two cartridges are used in each spatial template. The first is attached to a stationary mount inside a sealed enclosure. The second is installed on a six-axis mount, which extends into the enclosure through a flexible seal and allows us to adjust the width of the discharge gap and the relative lateral alignment between the two electrodes.

In the experiments presented here, we fill the enclosure with argon gas near atmospheric pressure and maintain a slight overpressure in order to flush out contaminants. The flushing manifold includes a set of nozzles directed into the discharge gap; this feature enables us to purge the gap frequently in order to maintain the highest purity in the most relevant region of the gas volume.

A 14-kHz driving voltage is applied across the water electrodes. An oscilloscope is used to display both the sinusoidal driving voltage (V) and a voltage ($V_C$) proportional to the charge stored by the "capacitor" made up of the electrodes, the dielectric slabs, and the gas gap. A comparison of these two signals provides important timing information: whenever an array of filaments discharges across the gap, $V_C$ exhibits a sharp jump at a distinct temporal phase of the driving oscillation. From this information we can measure both the number of discharge stages in each half-cycle and the time intervals between the stages. However, we do not have the ability to correlate particular spatial structures with specific temporal phases, as our camera acquires images with an integration time of 0.033 s.

### III. HEXAGONAL LATTICE

In [4] we used a single pin-grid to create a single-sided template for a square lattice. In this section we return to the theme of lattice potentials, but we focus on a two-sided, hexagonal-lattice template. The template is produced by a pair of pin-grid electrodes placed in a "mirror-image" arrangement, with each pair of opposed pins defining a lattice site. The sites are located at the vertices of contiguous equilateral triangles with a side length of 5.08(7) mm.

Fig. 2 shows examples of patterns observed with this setup. The top pair of panels reveals that filament footprints grow as the gap width increases. For a narrow gap ($z_0 = 0.5$ mm in Fig. 2a), filaments can fill the lattice completely, because no footprint extends far enough to affect the discharge process at adjacent lattice sites. By contrast, for a wide gap ($z_0 = 3.3$ mm in Fig. 2b), footprints can reach out laterally and inhibit the formation of filaments at neighboring sites. In panel 2b, this effect produces a patterned region characterized by a hexagonal tile in which a large star-shaped footprint covers a central lattice site as well as the six nearest-neighbor sites, while also being flanked by six smaller filaments located at the second-nearest neighbor sites. On average, the pattern generates one large footprint and two small footprints for every nine lattice sites, so the net filling fraction (F) is 1/3. Fig. 2c demonstrates that different patterns may stabilize in separate domains in our system. In the top half of the image, another pattern appears with F = 1/3; however, in this case, all of the footprints are of similar size and have the same six-pointed star shape. We note that this pattern has a condensed-matter counterpart in a common surface phase formed by adatoms on hexagonal crystal substrates [15, 16].

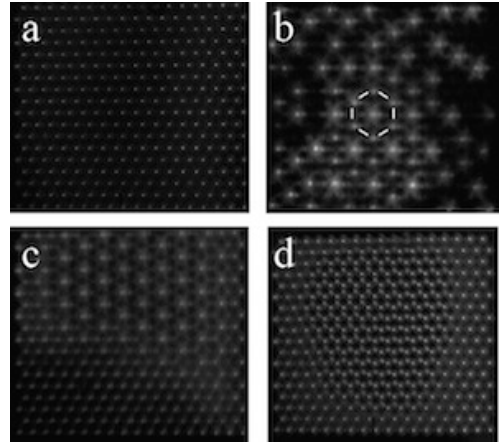

FIG. 2. Filament patterns in a hexagonal lattice template created by a pair of opposed pin-grid electrodes. (a) $V_0 = 1600$ V, $z_0 = 0.5(2)$ mm, one discharge stage; (b) $V_0 = 1700$ V, $z_0 = 3.3(2)$ mm, one stage; (c) $V_0 = 1960$ V, $z_0 = 2.5(2)$ mm, two stages; (d) $V_0 = 2660$ V, $z_0 = 1.0(2)$ mm, three stages. Further characteristics of the patterns are discussed in the text.

Whereas Figs. 2b and 2c show patterns that underfill the lattice, Fig. 2d provides an example of overfilling, which happens when filaments form in the regions between the sites of an already filled lattice. Overfilling occurs when $z_0$ is small and $V_0$ is sufficiently large. Under these conditions, filaments located at the lattice sites have small footprints that do not wield a long-range inhibitory effect. After these filaments ignite during an early phase of the rising (or falling) portion of driving voltage, the electric field in the gas can continue to increase and may cause breakdown at interstitial locations. It is important to note that our setup cannot detect all of the spatiotemporal structure of the overfilled patterns. While our electrical measurements reveal the timing of aggregated filament sub-arrays (or discharge stages), we cannot detect the temporal phase of each specific filament or sub-array. Thus, we do not have the full picture of how the observed spatial patterns are formed over time. Reconstructing the temporal sequence of sub-patterns would require methods comparable to those of [12].

## IV. OPPOSED SQUARE-LATTICE PIN-GRIDS WITH LATERAL MISALIGNMENT

The next investigation involves a double-sided square-lattice template produced by opposing pin-grid electrodes, one of which can be moved laterally relative to the other. The goal is to modify the anisotropic interactions between filaments and to observe the patterns that result. In this context, we say that a pairwise interaction is anisotropic if it depends not only on the distance between two filaments but also on the direction of their relative position vector.

There are two plausible mechanisms that might produce anisotropy when our opposing pin-grids are displaced laterally. First, if filaments are to terminate at the pin locations, they must tilt toward the lateral plane, as illustrated in Fig. 3a. Under these conditions, one might hope to model the system in terms of anisotropic interactions between "tilted dipoles" and, as suggested in [4], to develop an analogy for dipolar atoms and molecules in optical lattices [17, 18]. Second, if the lateral displacement alters the shape of filament footprints (Fig. 3b), then it is likely also to alter the footprint-dependent interactions. Anisotropic interactions can be expected to vary as footprints are stretched along the direction of displacement.

Our observations indicate that the effect of footprint-shape dominates over that of filament-tilt. Filaments are never seen to tilt more than 10° away from the normal to the dielectric surfaces. Instead, they strike across the gap along (or close to) the normal direction, and their footprints stretch along the dielectric surfaces toward the laterally displaced pins.

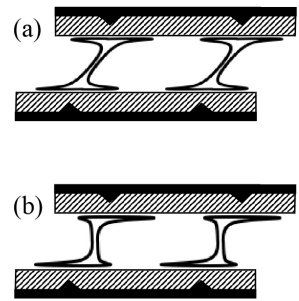

FIG. 3. Two scenarios leading to anisotropic interactions between filaments. The electrodes are indicated in black, the dielectric barriers are shaded, and the filaments are shown in white with black outlining. (a) Tilted filaments with undistorted footprints. It was suggested in Ref. 4 that this configuration could be modeled as an array of tilted dipoles, for which the anisotropic interaction is known. (b) "Normally directed" filaments with stretched footprints. This second scenario best represents what is usually observed in our experiments.

The anisotropic effects of stretched footprints are evident as we adjust the gap width and the lateral alignment of the electrodes. Using $x_0$ and $y_0$ to denote the values of the lateral shifts in the horizontal and vertical directions, and taking $z_0$ to be the value of the width of the gas gap, we designate the position of the movable electrode as $(x_0, y_0, z_0)$. Fig. 4 shows several examples of patterns that appear to stabilize preferentially at various times as the position is adjusted. We emphasize that the spectrum of patterns is not perfectly repeatable. In other words, there is not a one-to-one correspondence between the position setting and the pattern that the system selects. Rather, we conclude only that the selection probability for a given pattern may be enhanced over particular ranges of $(x_0, y_0, z_0)$.

Panels 4a-e show patterns characterized by filling fractions $F = 1/n$, with integer $n = 1, 2, …6$. While several of these patterns were observed in the studies presented in [4], the $n = 3, 5$, and 6 arrangements were rare. In our new setup, the $n = 3$ and 6 patterns remain relatively elusive and short-lived, regardless of the degree of lateral displacement. However, the $n = 5$ "knight's move" pattern is common and persistent. We also find that, even for relatively small displacements, it is uncommon for a single, unbroken pattern to cover the entire discharge plane. Most often, two or more patterns share the plane, as in 4c and 4d. Even when a single pattern is dominant (like the checkerboard in 4b) a fracture or dislocation is likely to appear.

Some additional conceptual discussion may help to make sense of these observations. In situations involving lateral displacement, one can define a lattice site as the lateral position of the midpoint between the tips of displaced, opposing electrode pins. Under this definition, each of the filaments sketched in Figs. 3a and 3b can be said to occupy a lattice site. However, interactions between adjacent filaments arise mainly from processes that occur not at these sites but rather at the edges of the filament footprints and near the dielectric surfaces. In order for a pattern to stabilize, these peripheral interactions must allow the footprints on each dielectric to produce a consistent "tiling" of the underlying pin-grid. An additional requirement of pattern formation is that the two tilings must be commensurable with each other. If each footprint on the front plate is connected by a single filament to a corresponding footprint on the back plate, then there must be a one-to-one correspondence between the front and back patterns.

The dependence of pattern formation on footprint shape is considered in Fig. 5. Panels 5a-e show patterns with the filling fractions represented in Figs. 4b-e. Let us focus momentarily on the $F = 1/4$ pattern in 5c, which is very common in our experiments. In this arrangement, each footprint "owns" three horizontally adjacent electrode pins and "shares" two additional pins with the footprints just above and below it. Our terminology here is based on the following provisional definitions of "owning" and "sharing:" a particular filament's footprint is said to own a specific pin if that filament is the only source of an ionization front that passes over the pin and deposits charge on the dielectric surface; if there are multiple such sources, of which the filament is just one, then the filament's footprint is said to share the pin.

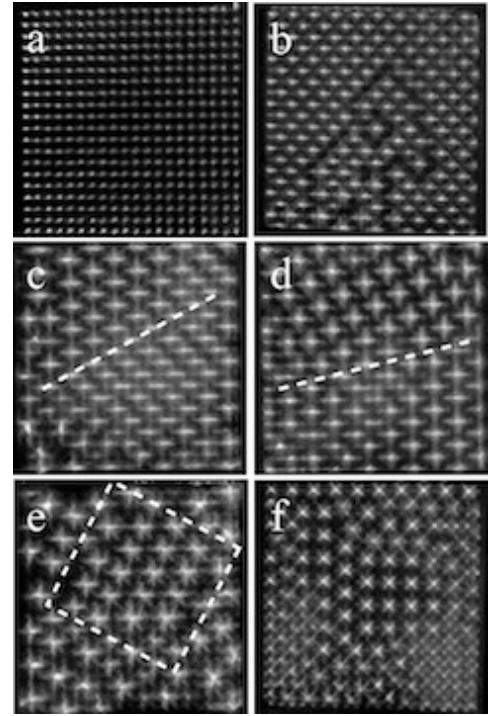

FIG. 4. Filament patterns for a double-sided square-lattice template with a relative lateral displacement between opposing pin-grid electrodes. The lattice period in the horizontal (x) and vertical (y) directions is 5.08 mm. The electrodes are laterally aligned when $x_0 = 0$ and $y_0 = 0$. The gap width is denoted as $z_0$ and the filling fraction as F. (a) F = 1; $V_0$ = 2440 V; $(x_0, y_0, z_0)$ = (0, 0, 3.1 mm). (b) F = 1/2, ; $V_0$ = 2570 V; (1.3 mm, 0, 2.1 mm). (c) F = 1/3 (bottom) and F = 1/4 (top); $V_0$ = 2470 V; (0.6 mm, 0, 3.8 mm). (d) F = 1/4 (bottom) and F = 1/5 (top); $V_0$ = 2400 V; (1.7 mm, 0, 3.8 mm). (e) F = 1/6; $V_0$ = 2500 V; (3.3 mm, 1.3 mm, 3.8 mm). (f) F = 1/4; $V_0$ = 2370 V; (0.25 mm, 0.25 mm, 1.8 mm), maximal lateral x and y offsets of one-half the lattice period.

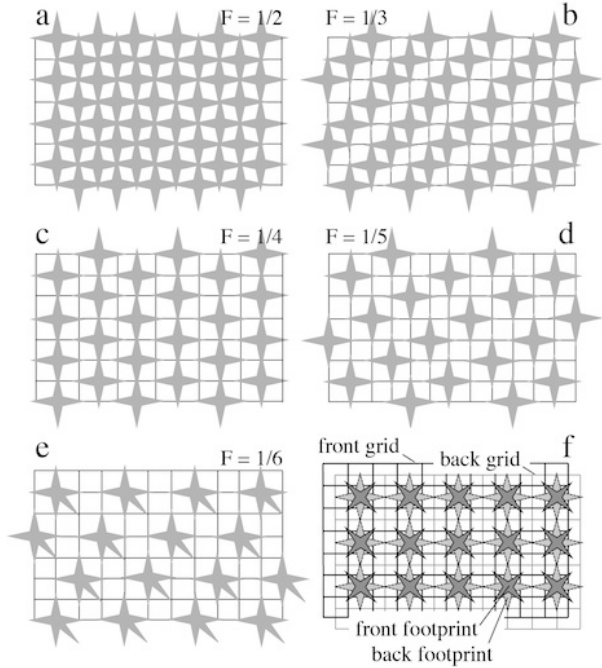

FIG. 5. Illustrations of some filling scenarios for the square lattice. Panels 5a-e assume no lateral displacement between the pin-grids. By contrast, panel 5f shows the "maximal offset" condition, with half-period displacements in the horizontal and vertical directions. (a) F = 1/2, as observed in Fig. 4b. (b) F = 1/3, comparable to the pattern displayed in the lower right portion of Fig. 5c. (c) F = 1/4, also seen in Figs. 4c and 4d. (d) F = 1/5, the "knight's move" pattern appearing in the top half of Fig. 4d. (c) F = 1/6, similar to the arrangement seen in the lower-left region of Fig. 4e. (d) F = 1/4 with a maximal lateral offset, as in Fig. 4f.

We conjecture that the most stable pattern for any given F requires every electrode pin to be either owned or shared. In other words, we expect the lateral position of each pin to be a site of surface-charge transfer during filament ignition. If any pin were to remain uninvolved in the discharge process, it would produce an unattenuated local field that could lead to instabilities. This suggested requirement is satisfied by all of the patterns illustrated in Figs. 5a-e, and it appears to be met in the patterns detected in Figs. 4a-e. Notably, in the F = 1/n patterns of Figs 4a, 5d, and 5e, each footprint owns n pins and shares none.

There is still an open question about how the lateral displacement affects pattern formation in these experiments. On one hand, the displacement may facilitate stabilization by breaking symmetries and reducing competition between patterns that have complementary orientations. For instance, the system may be altered to favor one of the two possible orientations of the n = 5 knight's move pattern. Or, for the F = 1/6 configuration, the displacement may favor the formation of footprint tendrils along the lower right diagonal (as in Fig. 5e) rather than along the upper left.

On the other hand, introducing a lateral shift may frustrate pattern formation by creating incommensurable conditions on the two sides of the discharge gap. The possibility of such an asymmetry is easiest to see when the lateral displacement is one-half of a lattice period along both x and y directions—the condition of maximal lateral offset. In this situation, represented in Figs. 4f and 5f, each footprint on the back plate owns a central pin and reaches out laterally to share its four closest neighboring pins. On the front plate, each footprint is centered at an interstitial location and stretches diagonally to take full ownership of the four nearest pins. If we take each shared pin to be "half owned," then the back footprint owns only three pins, while the front footprint owns four.

As can be seen in Fig. 5f, one-quarter of the pins on the back plate have no owner, so there is clearly a "pin-count asymmetry" between the front and back patterns. This fact provides one possible explanation for why the patterns in Fig. 4f stabilize only briefly and over relatively small domains. We note, however, that filaments in this situation tend to rearrange themselves intermittently in the two F = 1/4 geometries sketched in Figs. 5c and 5f, without seeming to prefer either of these arrangements to the other. Given that the arrangement in 5c eliminates the pin-count asymmetry between the front and back footprints, we suspect that the "shape asymmetry" between these footprints—which holds in either geometry—plays a more significant role in destabilizing the patterns.

## V. STRIPE TEMPLATE: A "SMECTIC" SYSTEM

We have just established that complex dynamics may arise when individual filaments terminate in opposing footprints with dissimilar shapes. In this section we implement an extreme case of this "end-to-end" filament asymmetry. The result is a system exhibiting a crossover between two distinct dynamical regimes, each rooted in a specific footprint geometry.

The experiments employ one flat, featureless electrode and one electrode made up of a series of straight ridges with a regular spacing, s = 5.08 mm. Each ridge has a sharp "knife edge" that produces a stripe of enhanced electric field in the gas outside the dielectric barrier. Because of the concentrated field, filament footprints on the side of the stripe electrode are elongated along the ridges and narrow in the transverse direction, much like the footprints observed in 1D DBD systems [8, 10]. The transverse confinement of surface charge results in a constraint on filament motion: every filament is tightly tethered to a particular stripe and can move only along the direction of that stripe.

On the side of the featureless electrode, the situation is very different. The footprints here are approximately disk-like, with characteristic diameter d. It will be key to this investigation that d is a controllable variable, since it will determine the coupling between filaments on adjacent stripes in the system. As we mentioned in the discussion of Figs. 2a and 2b, an increase in the discharge gap width leads to a corresponding increase in d, presumably because additional charge is produced in the longer filament channel and must be spread over an expanded surface area. Thus, we can adjust $z_0$ in order control d.

The images in Fig. 6 summarize our findings. For small gap widths, d is much smaller than s. Consequently, filaments lying along one stripe have negligible interactions with those lying along adjacent stripes. This transverse decoupling leaves each filament to interact only with its nearest collinear neighbors. Therefore, each stripe can be regarded as an independent 1D system. This is the situation pictured in Fig. 6a. Since filaments (or strings of filaments) on adjacent stripes can slide past each other easily, the entire system is analogous to a smectic liquid crystal, in which parallel molecular layers move independently in planes perpendicular to the stacking direction [19]. Of course, this analogy involves a reduction in dimensionality—from a 3D liquid to a 2D planar discharge. It also requires an unusual orientation for the liquid crystal's "director," which now lies along the filament axis and is therefore perpendicular to the stacking direction.

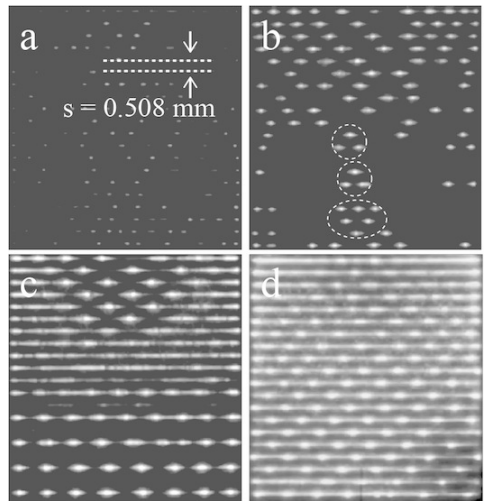

FIG. 6. Filament arrangements for the single-sided stripe template. Each image in panels a, b, and c corresponds to a discharge that occurs in a single temporal stage; panel d shows a two-stage discharge. (a) $V_0$ = 1400 V, $z_0$ = 0.5 mm. Filaments interact only with their collinear neighbors and move freely past the filaments on adjacent stripes. (b) $V_0$ = 1550 V, $z_0$ = 1.0 mm. Filaments can interact transversely and form short-range clusters (indicated in dashed circles). (c) $V_0$ = 1800 V, $z_0$ = 1.5 mm. Two patterned domains are separated by a disordered region. In the lower domain, breakdown is suppressed on alternate stripes. (d) $V_0$ = 2200 V, $z_0$ = 2.5 mm. The discharge is strongly driven, and the relatively large $z_0$ produces footprints with considerable reach. The resulting pattern displays collective horizontal motions that include phonon-like excitations.

Starting from the smectic condition (d < s), we adjust the gap width to introduce a tunable transverse coupling between filaments on adjacent stripes. Once the gap width is sufficiently large, a filament may interact with its nearest neighbors in the transverse as well as the longitudinal direction. The resulting 2D pattern depends to some degree on the voltage amplitude $V_0$. For small $V_0$, two filaments on adjacent stripes may bind together to form a stable pair. As $V_0$ is increased, larger clusters of filaments may form, each supporting a complex set of horizontal vibrational modes (Fig. 6b). Under certain conditions, discharges on one stripe may be completely inhibited because of the transverse effects of filaments on adjacent stripes (Fig. 6c). For high $V_0$, each stripe of the system becomes filled with a string of regularly spaced filaments, and the transverse interlocking of the strings produces a 2D crystal structure (Fig. 6d). While filaments still cannot move transversely in this last pattern, they can participate in large-scale collective motion along the stripe dimension. Such motion includes phonon-like excitations that propagate horizontally through the system.

Throughout these experiments, transitions analogous to freezing occur as the gap width is increased and d approaches s. To characterize the transition more precisely, we adopt terminology used elsewhere to distinguish the phases of colloids subjected to 1D periodic optical potentials [20, 21]. In this classification scheme, the term "locked smectic" is used for configurations like the one shown in Fig. 6b, which exhibit only short-range correlations between filaments on different stripes. The term "locked floating solid" applies to patterns such as that of Fig. 6d, which involve longer-range correlations and greater resistance to shear deformations. Using this terminology, we can identify two kinds of changes that are observable in our system, namely a "smectic → locked smectic" transition and a "locked smectic → locked floating solid" transition.

## VI. OPPOSED RING ELECTRODES

In [8] and [10], we examined 1D patterns that formed between two straight stripe electrodes. The boundary conditions were rather soft in those experiments, because it was impossible to establish a "hard edge" in the electric field profile produced by the electrode segments. In this next investigation we eliminate all end effects by using ring-shaped electrodes and imposing a cyclic boundary condition. Two electrodes are machined in the same fashion as was the stripe electrode of section V, except that in this case each ridge forms a circle with a diameter of 2.54 cm. The rings are placed in mirror-symmetric opposition. When a discharge filament forms between them, both of its footprints are elongated and follow the underlying circular contour of the machined ridges.

Our aim is to study pattern formation on the circle while treating the gap width $z_0$ and voltage amplitude $V_0$ as control parameters. Either of these parameters can be used to alter the amplitude of the applied electric field and the size of the charge footprint, which are the dynamically relevant variables. Fig. 7 shows a series of patterns obtained as the field amplitude is increased. As we perform these experiments, the number of filaments (N) in the system jumps intermittently, with a brief interval of disorder during each transition. When the field amplitude reaches a critical value, a second temporal discharge stage appears in each half cycle of the driving oscillation. The additional stage becomes evident in the $V_C$ trace on our oscilloscope and introduces new features in the filament patterns.

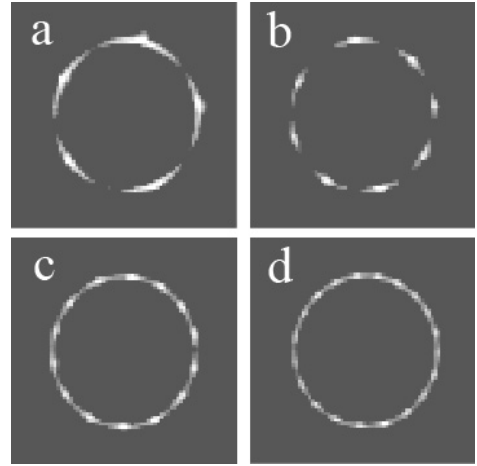

FIG. 7. Patterns on a double-sided circular template (diameter = 2.54 cm). (a) $V_0$ = 2440 V, $z_0$ = 2.9 mm, N = 5. (b) $V_0$ = 2250 V, $z_0$ = 2.1 mm, N = 8. (c) $V_0$ = 2220 V, $z_0$ = 2.0 mm, N = 14. (d) $V_0$ = 2150 V, $z_0$ = 1.6 mm, N = 20.

While this system is similar in several ways to the linear setup of [8], its behavior is novel in important respects. For example, the boundary conditions of the linear geometry allowed the distance between filaments to change continuously while N remained constant. Continuous change of this sort is forbidden in the present experiment, at least if filaments are to remain evenly distributed around the circle. Additionally, while the onset of a second discharge stage caused N to double in the linear system, it produces a less predictable (usually smaller) increase for the circular patterns. Nevertheless, N always takes on even values if there are two discharge stages during each voltage half-cycle. This effect is evident in Fig. 8.

The onset of the second discharge stage also initiates a change in the shape of filament footprints. The resulting pattern is illustrated in Fig. 9. On each dielectric surface there is an alternating sequence of two different footprint shapes: one is the familiar elongated shape common to single-stage discharges; the other is a narrow strip with "split ends." We refer to the overall arrangement as an "ex-dash" pattern because of its physical appearance (x—x—x—). Remarkably, each filament in the pattern terminates in an "ex" at one end and a "dash" at the other. This simple fact allows us to address the question of why N must be even under the two-stage discharge scenario. A stable circular pattern requires an integer number of "ex-dash" pairs, and this means that the total number of filaments must be a multiple of two. Of course, the dynamical cause of the "ex" footprint requires further investigation. The timing of filament ignition in the overall pattern also remains an open question.

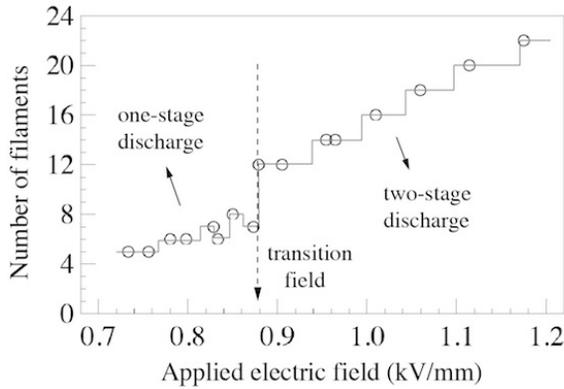

FIG. 8. The number of filaments on the circle (N) vs. applied electric field amplitude ($E_0$). The data are obtained as the gap width is increased from 1.5 mm to 3.0 mm, with the voltage amplitude ($V_0$) held constant. The value of $E_0$ is estimated as described in Section II; its uncertainty is contained within the circular symbol. The transition from a one- to two-stage discharge takes place near $E_0 = 0.88$ kV/mm.

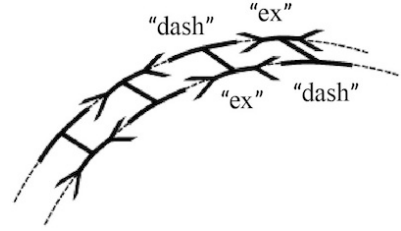

FIG. 9. Illustration of the "ex-dash" pattern, which is named for its appearance (x—x—x—), and which characterizes all of the stable configurations observed above the transition field indicated in Fig. 8. In order for this pattern to stabilize around the circle, the total number of filaments must be even.

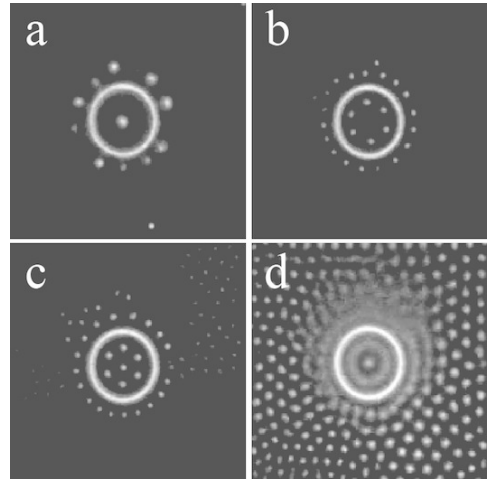

FIG. 10. Patterns produced with a disordered discharge on the ring. The top panels correspond to one-stage discharges, and the lower panels to two-stage discharges. (a) $V_0 = 2220$ V, $z_0 = 1.8$ mm. (b) $V_0 = 2540$ V, $z_0 = 1.8$ mm. (c) $V_0 = 2700$ V, $z_0 = 1.8$ mm. (d) $V_0 = 3100$ V, $z_0 = 1.8$ mm.

We conclude this section by noting the behavior of the system for field amplitudes higher than the values chosen for Fig 8. Under such conditions the filaments on the circle become disorganized, so that the ring is illuminated nearly uniformly. This behavior is consistent with that described in [8] and [10] for the 1D DBD. However, as indicated in Fig. 10, the over-driven ring system spawns filaments both inside and outside the circle. A small number of interior filaments may form a simple cluster, the stability of which appears to depend on the boundary condition established by the discharging gas around the circle. In this situation the circular discharge serves as a "filament corral," which might be seen as a classical analogue to the quantum corrals reviewed in [22].

The behavior of filaments spawned outside the discharging circle is also noteworthy. Over wide ranges of $V_0$ and $z_0$, these filaments are ejected in the form of bound pairs, triplets, or more complicated clusters [9]. However, over a small range of $V_0$ just above the value required for destabilizing the circular filament pattern, the bonding tendency appears to be weakened, and filaments move as "singlets" through the featureless outer region of the discharge plane. Single filaments may occasionally bounce off of one another, but they do not bind together as they often do for higher values of $V_0$. Moreover, when the plane is densely populated with filaments, the transition from bonding to non-bonding behavior becomes evident in a decrease in spatial regularity (the filaments move more like particles in a fluid than like particles in a crystal). We suspect that this "melting" phenomenon is a manifestation of the same pairwise interactions that produced the "smectic → locked smectic" transition described in section V.

## VII. PAIRING AND CLUSTERING INTERACTIONS

In this final investigation we generate stationary "pinned" filaments and study how they interact with nearby "free" filaments. The setup employs one featureless electrode and another electrode that is flat except for a few, isolated, conical pins (like those in sections III and IV but more sparsely distributed). This combination produces a single-sided template with point-like enhancements of the electric field on one side of the discharge gap. Because of the enhanced field, filaments form first at the pin sites. The footprints of these filaments tend to be disk-like, with characteristic radii that are only slightly smaller on the pin-side of the gap than on the featureless side. The pinned filaments in this system thus appear to behave like normal filaments in most respects, except for their inability to move laterally in the discharge plane.

While relatively low-amplitude driving voltages are capable of producing the pinned filaments, higher values of $V_0$ are required in order to generate additional, unpinned filaments in the flat regions of the template. These free filaments move laterally in the plane and occasionally interact with their pinned counterparts. The interactions involve three kinds of behavior. A free filament may (1) bounce off of a pinned filament, (2) become bound to a pinned filament, or (3) be annihilated by a pinned filament. By eliminating drift, our pinning technique allows us to study these behaviors over extended periods of time.

For gap widths between 0.5 mm and 2.5 mm, the pairwise bonding interaction exhibits a complicated dependence on the amplitude of the applied voltage (Figs. 11 and 12). For relatively high values of $V_0$, satellite filaments become attached to the central filament at a separation distance $r_1$, which is typically 3.5-4.0 mm. This is the minimum separation regardless of the number of attached satellites. As we reduce $V_0$, the filaments start to disappear one by one, until only a single satellite remains. The $r_1$ separation then persists as $V_0$ is lowered further toward a critical value, $V_0^{(1)}$, at which point the satellite may be extinguished, detached, or moved toward a new separation, $r_2$, in the neighborhood of 4.5-5.5 mm. In this last scenario, only a modest voltage attenuation $V_0^{(1)} \rightarrow V_0^{(2)}$ is required in order to initiate a significant spatial extension $r_1 \rightarrow r_2$. (We have not measured the functional dependence of $\Delta r$ on $\Delta V_0$ during this transition.) Subsequently, as we continue to reduce the amplitude, the two filaments remain separated by distance $r_2$ until $V_0$ reaches another critical value, $V_0^{(3)}$, whereupon the satellite is either extinguished or becomes detached and wanders away from the pinned filament.

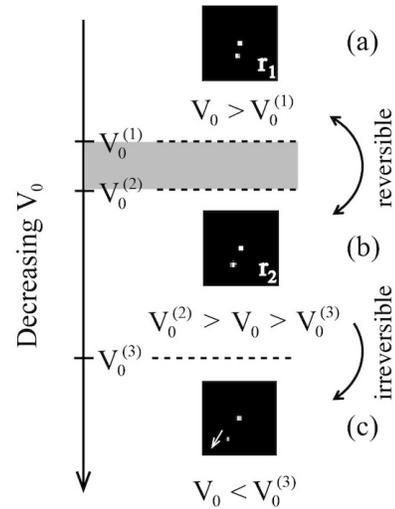

FIG. 11. Schematic diagram of the pairwise bonding interaction. (a) For $V_0 > V_0^{(1)}$, the satellite filament stays at a fixed distance $r_1$ from the pinned filament. (b) As $V_0$ is lowered to $V_0^{(2)}$, the satellite moves abruptly to a greater distance, $r_2$. The satellite can be moved back and forth between $r_1$ and $r_2$ by way of slight voltage adjustments. (c) If $V_0$ drops below the critical value $V_0^{(3)}$, detachment occurs.

In the course of these experiments, it is often possible to move the unpinned filament back and forth repeatedly between $r_1$ and $r_2$, simply by varying $V_0$ between high values ($> V_0^{(1)}$) and low values ($< V_0^{(2)}$). This "bistable" behavior is most clearly seen when the gap width lies within the range 0.5 mm $< z_0 <$ 1.5 mm, in which case the filaments have visible footprints less than 1 mm in radius.

A dynamical explanation for the stable (and bistable) bonding of filaments has yet to be developed, but we wish to note Stollenwerk's observations of intermittent filament pairing in a helium DBD. In [14] it is reported that, as two filaments approach each other, they form a quasi-stable bond at a separation of about 3 mm. In the same experiment, surface-charge measurements reveal that the distribution of negative charge around each filament exhibits a minimum near a radius of 1.5 mm. Thus, stability occurs when the filament separation allows the minima of the two surface-charge distributions to coincide. Given this result, it is not unreasonable to guess that bonding might generally require a precise geometric overlap between spatially commensurate charge distributions. Of course, if this idea were to apply to the $r_1 \leftrightarrow r_2$ transition in our system, two distinct overlap criteria would be required. As far as we know, such a situation has not been ruled out by any results from previous work.

The pairing phenomenon examined above is a limiting case of filament clustering. Clusters involving more than two filaments are seen frequently in conventional DBD systems [9], but they are rarely stable enough to allow for extended observation. In our final experimental configuration, we implement a "pinned-pair" technique to isolate two filaments in close proximity to each other, so that we can study their interactions with each other and with nearby free filaments. The main panel of Fig. 13 shows the array of filaments with which we begin, each located at a single electrode pin in a single-sided template. The narrow panels indicate how additional, free filaments can bond to the pinned pairs and create triangular or rhomboid clusters. The bond distances of the satellite filaments clearly depend on the baseline separations of the pinned pairs. This effect can be thought of in terms of the normal-mode oscillations that commonly occur in unpinned clusters: by pinning down one pair of filaments, we obtain an image resembling a snapshot of an elastic object (an unpinned cluster) at one phase of a complex oscillation. Additionally, it is relevant to note that the bonds tend to stretch/contract somewhat if we decrease/increase $V_0$ slightly. This effect could be the source of a "bulk property" analogous to compressibility, which may be of interest in future studies of filament clusters and patterns.

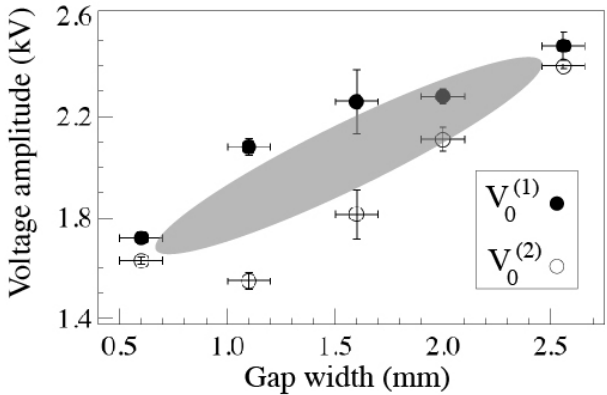

FIG. 12. Graph showing the range of parameters for which the bonding transition occurs. For each gap-width setting, the bond distance changes from $r_1$ to $r_2$ as $V_0$ is lowered from $V_0^{(1)}$ (filled circles) to $V_0^{(2)}$ (open circles). The change is reversible, as long as the satellite filament survives and $V_0$ does not drop too far below $V_0^{(2)}$. The transition occurs over the shaded region. Uncertainties reflect the variability of our measurements.

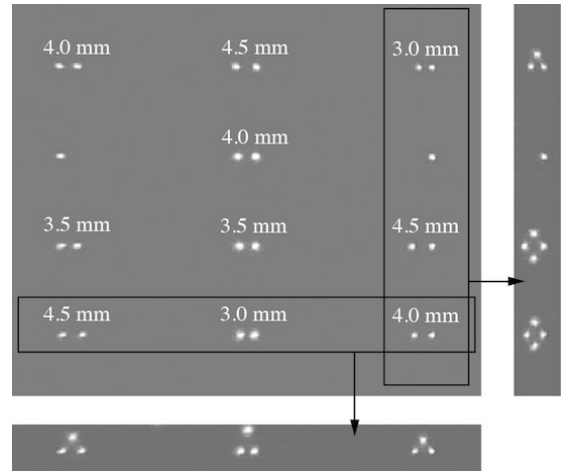

FIG. 13. Single-sided template for "pinned-pair" studies of clustering and inhibition phenomena. The main panel shows filaments localized at electrode pins. Separations of paired pins are indicated. Clusters form when free (unpinned) filaments attach to the pinned filament pairs. Examples of stable clusters appear in the narrow panels.

The pinned filaments shown in Fig. 13 can be used also to study the processes of annihilation and inhibition, which occur when one filament frustrates the formation of another. Once again, the footprint radius is the relevant variable, since it sets the spatial scale of pairwise interactions. In this experiment we change the footprint radius by varying the gap width $z_0$ and keeping $V_0$ constant. Starting with only pinned filaments, and with each pinned pair having a different baseline separation, we increase $z_0$ and record its value whenever a filament is annihilated. Annihilation always occurs within the remaining filament pair that has shortest separation. After each of the pairs has been reduced to a single filament, we reverse the procedure and decrease $z_0$, recording its value whenever a pair is reestablished. Re-ignition always occurs for the pair of pins with the largest separation. For each scan of $z_0$, the four baseline separations provide separate determinations of the annihilation and re-ignition distances. The graph in Fig. 14 shows that each of these distances has an approximately linear dependence on $z_0$. The obvious hysteresis can be explained in terms of the usual DBD memory effect, which tends to sustain either a single filament or a pair, even if a transition from one to the other is imminent. We leave it to future investigations to explore these annihilation and inhibition effects in greater detail.

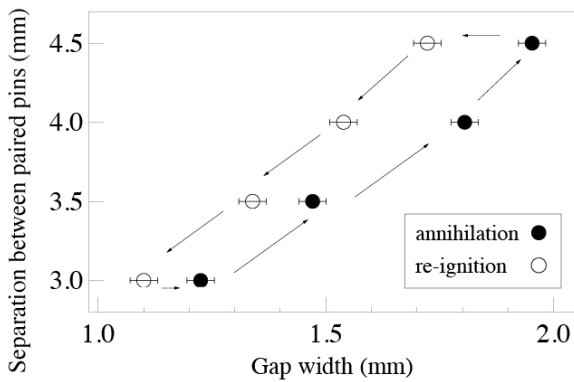

FIG. 14. Graph showing the onset of filament annihilation and re-ignition for four different pairs of pins. The lower left corner corresponds to full ignition at all pin positions. In the upper right corner, only one filament remains within each pair of pins. Uncertainties reflect the variability of our measurements.

## VIII. CONCLUSION

The experiments described above demonstrate the versatility of our method of designing, controlling, and studying DBD filament interactions and patterns. Transparent electrodes imprinted with spatial features can be used to create a wide range of spatial templates for the lateral arrangement of filaments. These arrangements may be of interest for several reasons. From the standpoint of basic research, they can lead to interesting connections with a variety of other pattern-forming systems. In addition, through filament-manipulation techniques like those of Section VII, it is possible to perform measurements with relevance to ongoing efforts to model the dynamics of DBD systems. In applied contexts, spatial templates could prove useful in the design of DBD actuators.

With regard to the specific templates examined in this paper, we must emphasize that certain questions remain unanswered and may be worth revisiting. Specifically, we have not addressed the timing of filament sub-arrays in any of our patterns. Achieving a full spatiotemporal characterization of our patterns would require an approach comparable to that of [12]. Also, owing to the fact that our imaging system is sensitive only to the optical footprints of discharge filaments, we have been unable to draw any strong conclusions relating to charge footprints. Thus, we have only hypothesized about the footprint models of Section IV and the pairwise interactions of Section VII. Our analysis could be tested through the use of charge-sensitive imaging techniques like those of [13] and [14].

We thank Phil Jasperse of the Calvin College machine shop for his contributions to the design and construction of our experimental system. We also thank Josiah Sinclair and Jonathan Shomsky, who helped with the production of several figures. This work was supported in part by NSF Grant No. PHY-1068078.

*mwalhout@calvin.edu